\documentclass[traditabstract]{aa}
\usepackage{graphicx}
\usepackage{txfonts}
\usepackage{natbib}
\usepackage{threeparttable}

\newcommand{\ltsim}{\protect\raisebox{-0.5ex}{$\:\stackrel{\textstyle <}{\sim}\:$}}

\begin{document} 
  \title{Polycyclic aromatic hydrocarbon feature deficit of starburst galaxies in the AKARI North Ecliptic Pole Deep Field}
\author{K.Murata
  \inst{1,2}
  \and H.Matsuhara\inst{1,2}
  \and H.Inami\inst{3}
  \and T.Wada\inst{1}
  \and T.Goto\inst{4}
  \and L.Armus \inst{5}
  \and C.Pearson\inst{6,7,8}
  \and S.Serjeant\inst{7}
  \and T.Miyaji\inst{9}
  %\fnmsep
  %\thanks{Just to show the usageof the elements in the author field}
}

\institute{
  Institute of Space and Astronautical Science, Japan Aerospace Exploration Agency, Sagamihara, 229-8510 Kanagawa, Japan\\
  \email{murata@ir.isas.jaxa.jp} 
  \and
  Department of Space and Astronautical Science, The Graduate University for Advanced Studies, Japan
  \and
  National Optical Astronomy Observatory, Tucson, AZ 85719, USA
  \and
Institute of Astronomy and Department of Physics,National Tsing Hua Universit
y, No. 101, Section 2, Kuang-Fu Road, Hsinchu 30013, Taiwan, R.O.C
%Institute of Astronomy and Department of Physics, National Tsing Hua University, No. 101, Section 2, Kuang-Fu Road, Hsinchu 30013, Taiwan, R.O.C
  \and
  Spitzer Science Center, Calfornia Institute of Technology, MS 220-6, Pasadena, CA 91125
  \and
  RAL Space, STFC Rutherford Appleton Laboratory, Chilton, Didcot, Oxfordshire, OX11 0QX, UK
  \and
  Department of Physical Sciences, The Open University, Milton Keynes, MK7 6AA, UK
  \and
  Oxford Astrophysics, Oxford University, Keble Road, Oxford OX1 3RH, UK  
  \and
  Instituto de Astronom\'{i}a, Universidad Nacional Aut\'{o}noma de M\'{e}xico, Ensenada, Baja California, Mexico
}

\date{Received March 2 2014; accepted April 4, 2014}
\abstract
    {We study the behaviour of polycyclic aromatic hydrocarbon emission in galaxies at $z=0.3-1.4$ using 1868 samples from the revised catalogue of AKARI North Ecliptic Pole Deep survey.
      The continuous filter coverage at 2-24$\mu$m makes it possible to measure 8$\mu$m luminosity, which is dominated by polycyclic aromatic hydrocarbon emission for galaxies at up to $z=2$.
      We compare the IR8 ($\equiv L_{IR}/L(8)$) and 8$\mu$m to 4.5$\mu$m luminosity ratio ($\nu L(8)/\nu L(4.5)$) with the starburstiness, $R_{\rm SB}$, defined as excess of specific star-formation rate over that of main-sequence galaxy.
      All AGN candidates were excluded from our sample using an SED fitting.
      We found $\nu L(8)/\nu L(4.5)$ increases with starburstiness at log $R_{\rm SB}$ $<$0.5 and stays constant at higher starburstiness.
      On the other hand, IR8 is constant at log $R_{\rm SB}$ $<$ 0, while it increases with starburstiness at log $R_{\rm SB}$ $>$0.
      This behaviour is seen in all redshift range of our study.
      These results indicate that starburst galaxies have deficient polycyclic aromatic hydrocarbon emission compared with main-sequence galaxies.
      We also find that galaxies with extremely high $\nu L(8)/\nu L(4.5)$ ratio have only moderate starburstiness.
      These results suggest that starburst galaxies have compact star-forming regions with intense radiation, which destroys PAHs and/or have dusty HII regions resulting in a lack of ionising photons.
    }
    \titlerunning {PAH deficit of starburst galaxies in AKARI NEP-Deep Field}
    \maketitle

\section{Introduction}
Polycyclic aromatic hydrocarbon (PAH) emission has received much attention in recent years because of its properties, which provide us with key parameters in galaxy evolution.
The PAHs are thought to be located in photo-dissociation regions (PDRs) and are regarded as star-formation rate tracers, since they are excited by UV light from young stars and emit their energy at 3.3, 6.2, 7.7, 8.6, and 11.3 $\mu$m.
Of these values, 7.7$\mu$m emission is the strongest and dominates at 8$\mu$m luminosity even in broad band filters.
Therefore, these emissions correlate with the infrared luminosity, $L_{IR}$ \cite[]{2007ApJ...660...97C}, which traces the dust-obscured starformation, since most of the energy is absorbed by dust and re-radiated at infrared. 
On the other hand, the PAHs also reflect the physical conditions of the inter stellar matter in galaxies.
If UV photons are absorbed by dust in HII regions and cannot excite the PAHs in PDRs, or PAHs are destroyed by harsh radiation from a strong starburst or active galactic nuclei (AGNs), the PAH emission would be lower compared with the infrared luminosity \cite[]{2008ARA&A..46..289T}.
Therefore, local ultra luminous infrared galaxies (ULIRGs;$L_{IR}$ $>$ 10$^{12}L_{\odot}$) that are very dusty and have compact star-forming regions and/or strong AGNs have a deficiency in PAH emission.
\par
However, the PAHs show different behaviour in a high-$z$ universe.
At $z$$\sim$0.5 and $z$$\sim$1, ULIRGs with a large rest-frame $L(8\mu$m$)/L(5\mu$m$)$ colour, which corresponds to large PAH equivalent width, were found, which must have an extended star-forming region \cite[]{2010A&A...514A...5T}.
The ULIRGs at higher $z$ were also observed to have a higher $L(8)$ at a given $L_{IR}$ than their local counterpart \cite[]{2008ApJ...675..262R,2009ApJ...700..183H}.
It leads an overestimation of infrared luminosity at $z>1.5$ when it is derived from only a 24$\mu$m band photometry with local SED templates, since the 8$\mu$m emission is redshifted into the 24$\mu$m band, which has been called as ``mid-infrared excess'' problem \cite[]{2007ApJ...670..156D}.
\par
Recently, this excess emission is explained with two modes of star formation, ``normal'' and ``starburst'' \cite[]{2011A&A...533A.119E}.
Galaxies with ``normal'' star-formation have a tight correlation between stellar mass and star formation rate, which is commonly referred to as the ``star-forming main sequence'' \cite[]{2007ApJ...660L..43N,2007A&A...468...33E} with an intrinsic scatter \cite[]{2012ApJ...754L..14S,2013ApJ...778...23G}.
%($\sim$0.3dex; Salmi et al.2012, Guo et al.2013).
``Starburst'' galaxies are defined with specific star-formation rate (sSFR) above the scatter, while ``passive'' are with sSFR under the scatter.
The sSFR of the main sequence increases with redshift, while the IR8 ($\equiv L_{IR}/\nu L(8)$) has a value of 4$\pm$1.6, which is independent of redshift \cite[]{2011A&A...533A.119E}.
The IR8 correlates with both a compactness in the star-forming region and a starburstiness that is defined by the excess of sSFR: $R_{\rm SB}\equiv$sSFR/sSFR$_{MS}$.
\cite{2012ApJ...745..182N} also showed that $L(8)$/$L_{IR}$(=1/IR8) decreases with an excess of sSFR at $z$=1 and $z$=2.
They showed that intensive quantities, which are averaged over the entire galaxy, such as the sSFR and SED shape, are more fundamental parameters in determining the physical conditions in PDRs than extensive quantities that are integrated over the entire galaxy, such as infrared luminosity and stellar mass.
However, owing to sparse filter sampling at 8-24$\mu$m in the Spitzer Space Satellite \cite[]{2004ApJS..154....1W}, they could not pay much attention to $z$$<$1.
Nonetheless, since the co-moving star-formation rate dramatically changed at this epoch \cite[]{2005ApJ...632..169L,2005ApJ...630...82P}, investigating the PAH behaviour of galaxies at the intermediate redshift is invaluable to explore galaxy evolution.
\par
In contrast, the Japanese led AKARI satellite \cite[]{2007PASJ...59S.369M} has a continuous filter coverage at 2-24 $\mu$m with nine photometric bands in the Infrared Camera \cite[]{2007PASJ...59S.401O}.
With these nine bands, a large and deep galaxy survey has been conducted towards the North Ecliptic Pole \cite[]{2006PASJ...58..673M,2008PASJ...60S.517W}, which makes it possible to measure 8$\mu$m luminosities at $z$=0.3-0.7, 0.7-1.2, and 1.2-1.4 using the S11, L15, and L18W bands without associated uncertainties from the K-correction.
This catalogue has been recently updated by \cite{2013A&A...559A.132M} and the detection limit is improved, as $\sim$60$\mu$Jy at S11 band.
\par
In this paper, we explore the PAH behaviour of galaxies at $z$=0.3-1.4 using the revised AKARI NEP-Deep catalogue.
In section 2, the data used in this study is described.
In section 3, we show that the starburst galaxies have a relative weakness of PAH emission with respect to infrared luminosity regardless of redshift.
In section 4, we discuss our results.
In section 5, the conclusion is given.
We adopt a cosmology with ($\Omega_m$, $\Omega_\Lambda$, $H_0$) = (0.3, 0.7, 70km s$^{-1}$ Mpc$^{-1}$).
An initial mass function of \cite{2003PASP..115..763C} is assumed.

\section{Data and sample}
\label{data}
In this study, we used the AKARI North Ecliptic Pole Deep survey revised catalogue (Murata et al.2013) to measure the PAH emission strength at $z$=0.3-1.4.
It has an advantage of covering 2-24 $\mu$m wavelengths continuously, which makes it possible to measure 8$\mu$m feature at up to $z$=2, where PAH emission at 7.7$\mu$m and thermal emission from a very small grain heated by AGNs are dominant sources.
A total of 1868 galaxies at $z$=0.3-1.4 with photometry in all nine bands were selected from the catalogue.
All photometry was performed with an aperture radius of 6.3 and 6.0 arcsec for NIR and MIR, with which flux calibration was conducted (Murata et al.2013).
The NIR and MIR sources are matched through the matching with a ground-based catalogue (u* to Ks bands, see below).
The 50\% completeness limits of the survey are as deep as 56, 55, 38, 29, 34, 50, 86, 94, and 247$\mu$Jy in the N2, N3, N4, S7, S9W, S11, L15, L18W, and L24 bands, respectively.
\par
To estimate the physical properties of the AKARI sources, we used ground-based data taken with CFHT/MegaCam and WIRCam, which covers most of the AKARI NEP-Deep field with eight $u^*g'r'i'z'YJKs$ bands \cite[]{2014arXiv1403.7934O}.
The $YJKs$ stacked image was used as a detection image when sources were extracted using SExtractor \cite[]{1996A&AS..117..393B} with the dual mode for all eight bands.
The colours were measured with a 2.0 arcsec aperture radius with the same area for all eight bands, while the total magnitudes were scaled with the $Ks$ mag measured with the $Mag\_Auto$.
The detection limit of this catalogue is $Ks\sim$22.5 mag.
The CFHT sources were cross-matched with the AKARI NIR and MIR catalogue with 1.5 and 2.5 arcsec search radii, respectively, which are slightly small compared with the positional accuracy to avoid a chance coincidence of the matching with different objects.
Some objects were unmatched with the CFHT catalogue due mainly to the field coverage and the detection limits.
\par
\begin{figure}[t]
\centering
\includegraphics[width=65mm,angle=270]{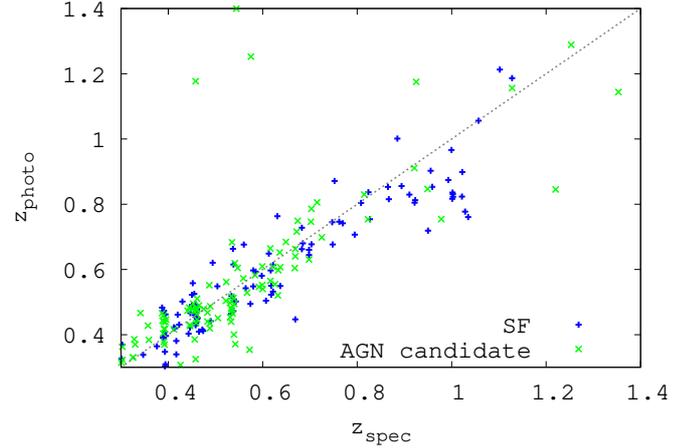}
\caption{\label{fig:zpzs}
Comparison between photometric and spectroscopic redshifts.
Blue pluses indicate star-forming galaxies, while green crosses indicate AGN candidates.
}
\end{figure}

We used spectroscopic redshifts when available; otherwise, we used photometric redshifts.
Among 1868 objects, 225 have a spectroscopic redshift, which were obtained by MMT/Hectospec, WIYN/Hydra \cite[]{2013ApJS..207...37S}, Keck/DEIMOS, and Subaru/FMOS (Appendix \ref{fmosobs}) with at least two emission or absorption line features.
We note that a fraction of high-z redshifts were obtained with [OII] line measured with the DEIMOS, while the others were determined with H$\alpha$ line by the FMOS.
Photometric redshifts were estimated using a publicly available code $LePHARE$ \cite[]{2006A&A...457..841I}, with which 62 galaxy templates from \cite{2007A&A...476..137A} were fitted to $u^*$ to N4 bands.
The systematic magnitude offsets for each band were calculated using the best fit templates of objects with spectroscopic redshift using the $AUTO\_ADOPT$ option.
The photometric redshifts were compared with the spectroscopic redshift in Fig.\ref{fig:zpzs}.
The normalised median absolute deviation was $\Delta z/(1+z)$=0.043 for star-forming galaxies (blue; see later paragraph for classification) and 0.046 for AGN candidate (green), while the outlier rates, defined as $|z_{photo}-z_{spec}|>0.15 (1+z_{spec})$, were 2.7 and 7.1\%, respectively.
\par
To estimate the PAH emission strength, we measured rest-frame 4.5$\mu$m and 8$\mu$m luminosities.
Since PAH emission dominates 8$\mu$m luminosity of star-forming galaxies while continuum emission dominates at 4.5$\mu$m luminosity, the 8$\mu$m to 4.5$\mu$m luminosity ratio roughly corresponds to an 8$\mu$m PAH equivalent width. 
Using the AKARI S7-L18W bands, these luminosities were calculated with less uncertainties from K-correction.
For example, the L15 flux was used for calculation of 8$\mu$m luminosities at $z=1$ using the following equation:
\begin{eqnarray}
[8]_{abs}=[L15]-DM (z=1) -(kcor(L15) + ([8]-[L15])_{abs}^{temp}),
\end{eqnarray}
where DM is the distance module and the templates from \cite{2006ApJ...642..673P,2007ApJ...663...81P} were applied. 
For 8$\mu$m luminosities at $z$=0.3-0.7, 0.7-1.2, and 1.2-1.4, we used S11, L15, and L18W bands, while we used S7, S9W, and S11 bands for 4.5$\mu$m luminosities.
The IRAC 2 and IRAC 4 bands were applied as 4.5$\mu$m and 8$\mu$m bands for the K-correction. 
\par
%Our samples were divided into three categories, star-forming (SF), AGN candidate and elliptical galaxies, through a $\chi^2$ SED fitting using the $LePHARE$.
Our samples were divided into three categories, star-forming (SF), AGN candidate and elliptical galaxies, for which a best $\chi^2$ fitting template was provided with the $LePHARE$.
The N2-L24 bands were fitted to 25 POLLETTA templates of which three are elliptical, 13 are star-forming, and nine are AGNs.
Among the AGN templates, three are type 1, three are type 2 or 1.8, two are composite (starburst+AGN), and one is a torus model, where the type 2 SEDs are dominated by host galaxies (Polletta et al.2007).
Hence, even galaxies which are not dominated by AGNs are identified as AGNs and we call these sample AGN candidate.
This conservative categorisation was performed to obtain a pure SF sample.
Throughout this paper, AGN candidates are used only for comparison, and our conclusion is based only on star-forming galaxies.
\par
To estimate the infrared luminosity more reliably, we also used Herschel/PACS 100$\mu$m and 160$\mu$m (Serjeant et al. in prep.) and SPIRE 250$\mu$m, 350$\mu$m, and 500$\mu$m data (Pearson et al. in prep.).
The survey with PACS 100 and 160$\mu$m bands covers most of the NEP-Deep region with 5$\sigma$ sensitivity of 8 and 16 mJy, while the survey with the SPIRE 250, 350, and 500$\mu$m bands covers nearly half of the field with 5$\sigma$ sensitivity of 26, 21.5, and 31 mJy and covers entire field with 45, 38, and 54 mJy.
%4.8mJy 9.6mJy fie 100, 160 3sigma (p4)
%OT2 SPIRE 9.0 7.5 10.8 1sigma (p41)
%GT2 SPIRE 5.2 4.3 6.2 1sigma (p41)
We cross-matched our sample with the PACS catalogue using a search radius of 5 arcsec and 160 objects were matched.
The search radius was determined as 2.5$\sigma$ of the source separation distribution.
The SPIRE photometry was made by fitting an elliptical Gaussian to the SPIRE timelines by assuming an initial position (RA, Dec) from the AKARI NEP catalogue.
As a result, 1068 out of 1868 objects have SPIRE photometry.
The errors of the absolute PACS photometry was estimated as large as $\sim$20\%, while the SPIRE $\sim$10\%.
\par
\begin{figure}[ht]
\centering
\includegraphics[width=65mm,angle=270]{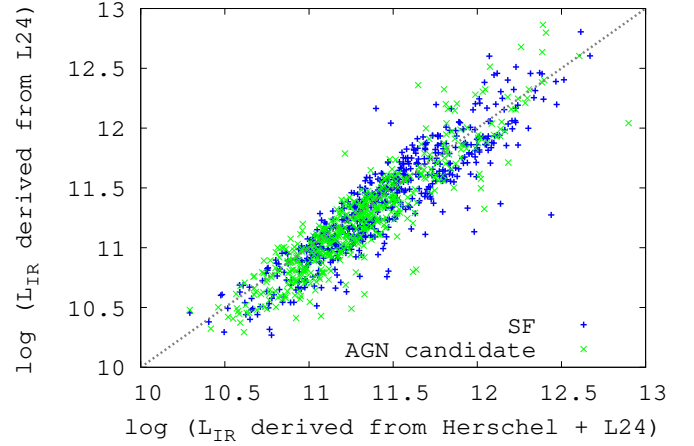}
\caption{\label{fig:lircom}
Comparison of infrared luminosities derived from only L24 and from L24 + Herschel bands.
Blue pluses indicate star-forming samples while green crosses indicate AGN candidates.
Grey dotted line indicates y=x line.
}
\end{figure}

The infrared luminosities (integrated over 8-1000$\mu$m) were estimated via an SED fitting with the \cite{2001ApJ...556..562C} SED library using the L24, 100 and 160$\mu$m (PACS), 250, 350, and 500$\mu$m (SPIRE) photometry.
We did not use AKARI N2 to L18W bands to avoid PAH emission which may overestimate the infrared luminosity.
For objects undetected by any Herschel band, infrared luminosities were derived only from L24 band.
To estimate the accuracy we compared the infrared luminosities derived from L24 and those from mid- to far-infrared bands, as shown in Fig.\ref{fig:lircom}.
They agreed within 0.2dex for both SF and AGN candidates, although fainter galaxies tend to have a lower luminosity when derived only from L24 bands.
It may result from the confusion limit of the SPIRE photometry, which can overestimate the flux, since we performed photometric measurements even if they are below the detection limit.
However, we found our conclusions to be insensitive to the inclusion or exclusion of the Herschel data, so we believe the confusion effect not to be serious. 
Our SEDs show no mid-infrared excess (see also Daddi et al. 2007), since the redshift range of our sample is $z=0.3-1.4$, where the infrared luminosities from the $24\,\mu$m photometry agree very will with those from far-infrared bands\cite[]{2010A&A...518L..29E}.
The infrared luminosities were converted into star-formation rate via multiplication of 1.09$\times$10$^{-10}$[SFR/L$_\odot$].
\par
\begin{figure}[ht]
\centering
\includegraphics[width=65mm,angle=270]{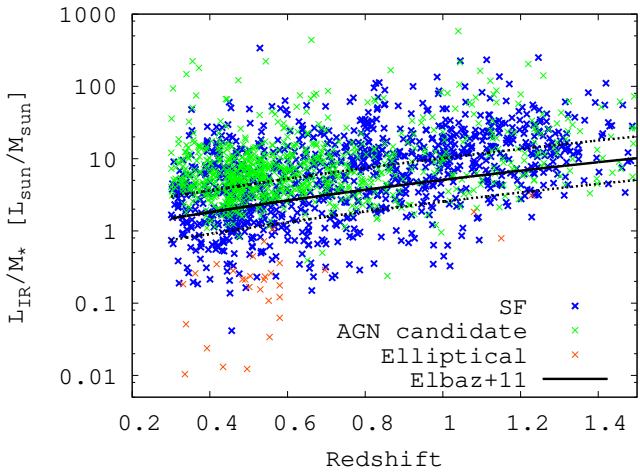}
\caption{\label{fig:zssfr}
Infrared luminosity divided by the stellar mass against redshift.
Blue, green, and orange points indicate star-forming, AGN candidates, elliptical galaxies, respectively.
Black lines correspond to the specific star-formation rate of the main-sequence galaxy with 0.3 scatter at given redshift defined by Elbaz et al. (2011).
}
\end{figure}

For each object, we calculated the specific star-formation rate, sSFR, defined as the star-formation rate divided by stellar mass.
Masses were estimated from a $\chi^2$ SED fitting using \cite{2003MNRAS.344.1000B} SED models with solar metallicity.
In this SED fitting, we used MegaCam $u^*g'r'i'z'$, WIRCam $YJKs$, and AKARI N2, N3, and N4 bands.
The typical mass of our sample is log $M_*$= 10.4, 10.5, and 10.6 for $z$=0.3-0.6, 0.6-0.9, and 0.9-1.4 bins, respectively.
Most of our samples have values of log $M_*/M_\odot$$>$10, which is similar to Nordon et al. (2012).
The starburstiness is defined as the deviation of sSFR from that of main-sequence, defined by \cite{2011A&A...533A.119E}.
\begin{eqnarray}
R_{SB} = sSFR/sSFR_{MS}\\
sSFR_{MS} [Gyr^{-1}] =26\times t^{-2.2}_{cosmic},
\end{eqnarray}
where $t_{cosmic}$ is the age of the Universe at that redshift.
We did not estimate the $\rm sSFR_{MS}$ from our sample because high-$z$ main-sequence galaxies were incomplete.
Fig.\ref{fig:zssfr} shows the variation of specific star-formation rate with redshift for our sample with a scatter of 0.3dex. 
We sample a wide range of the population, even below that of the main sequence at lower redshifts.
However, our MS sample at $z>$0.8 is incomplete due to a flux limit, which we discuss in the section \ref{discuss}.
The figure also shows AGN candidate galaxies have a higher $L_{IR}/M_*$ than star-forming galaxies while all elliptical galaxies have a lower $L_{IR}/M_*$ than the main-sequence.
The SFR and $R_{\rm SB}$ were derived from the infrared luminosity.
%We note that although $R_{\rm SB}$ of AGN candidates might be overestimated, this does not affect our conclusion because they are based only on the SF galaxies; the AGN candidates are only used for comparison.
We note that $R_{\rm SB}$ of AGN candidates might be overestimated. However, this does not affect our conclusions because they are based only on the SF galaxies; the AGN candidates are only used for comparison.

\section{Results}
Previous studies using the Spitzer Space Telescope (Elbaz et al. and Nordon et al.) showed that high starburstiness galaxies have lower PAH emission than expected from the infrared luminosity.
Elbaz et al. (2011) showed that IR8 ($\equiv L_{IR}/\nu L(8)$) correlates with $R_{\rm SB}$, where 8$\mu$m luminosities were measured with IRAC4, IRS 16$\mu$m, and MIPS 24$\mu$m bands for galaxies at $z=$ 0.3, 1, and 2, respectively.
Nordon et al. (2012) obtained similar results for $z\sim$1 and $z\sim$2 galaxies using ultra deep IRS spectra and showed that the spectra are dominated by PAH emission.
In our work, IR8 and $\nu L(8)/\nu L(4.5)$ ratio of galaxies at $z=$0.3-1.4 were obtained using the continuous wavelength coverage of the AKARI/IRC.
The AGNs and elliptical galaxies were separated from our sample to study the behaviour of PAHs in star-forming galaxies.
\begin{figure*}[ht]
\centering
\begin{minipage}{0.49\hsize}
\includegraphics[width=50mm,angle=270]{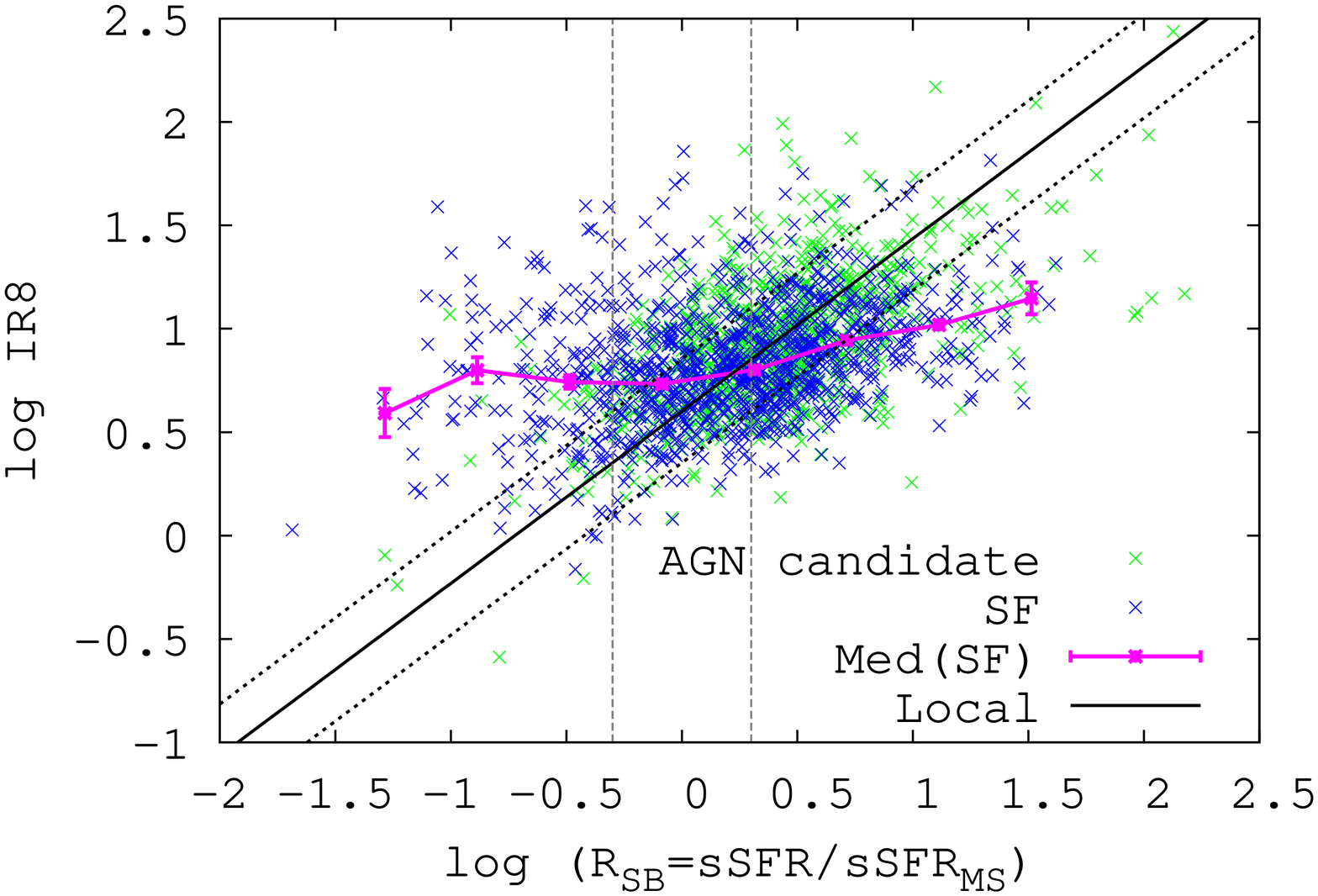}
\end{minipage}
\begin{minipage}{0.49\hsize}
\includegraphics[width=50mm,angle=270]{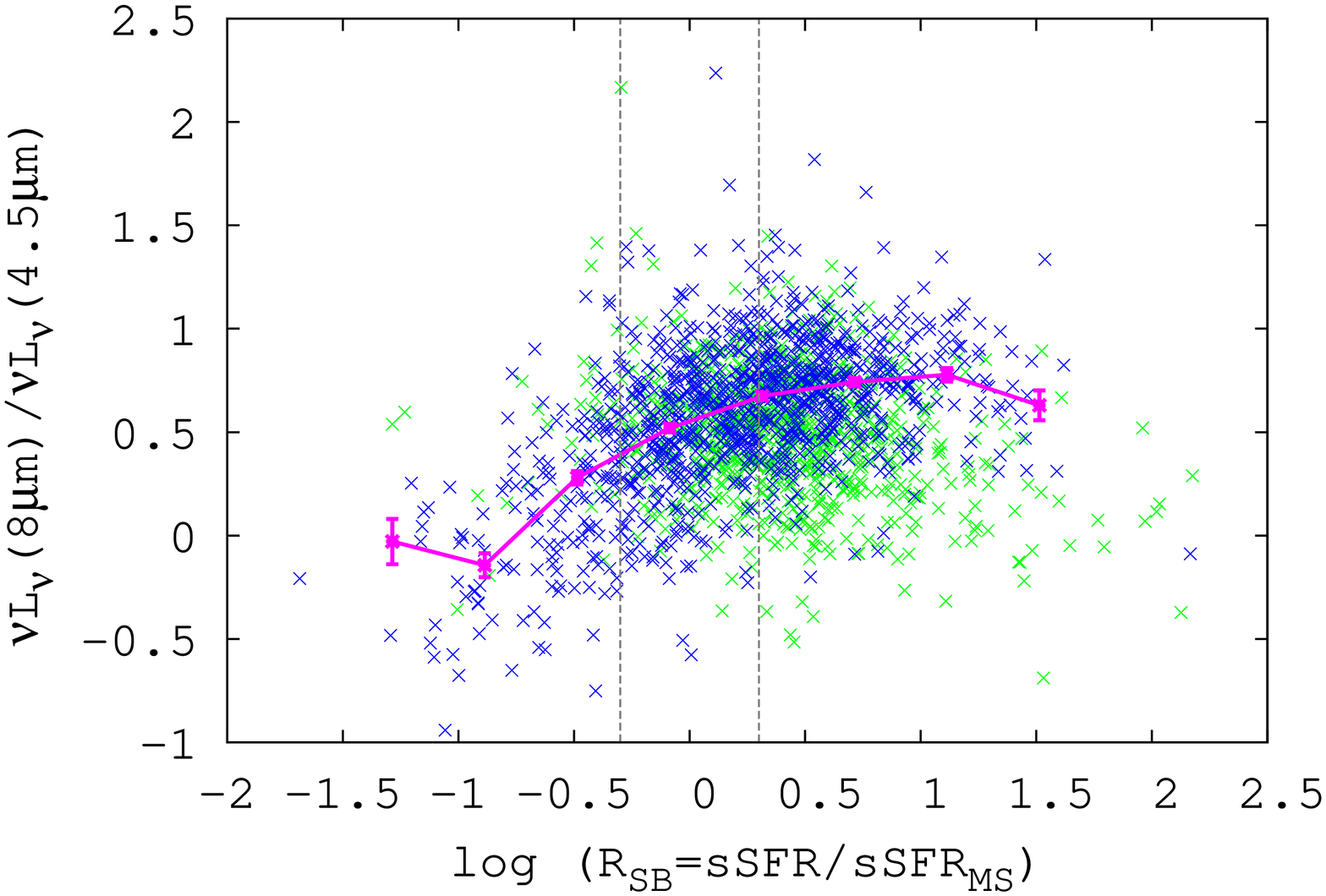}
\end{minipage}
\caption{
  Left: IR8 against starburstiness.
  Green and blue points indicate AGN candidate and star-forming sample.
  Magenta line indicates the median value of the IR8 for star-forming sample.
  Black lines show the local relation between IR8 and $R_{\rm SB}$ and its scatter from Elbaz et al. (2011).
  Grey dotted vertical lines shows the extent of the main-sequence.
  Right:  $\nu L(8)/\nu L(4.5)$ ratio against starburstiness.
  Symbols are the same as the left panel.
\label{fig:each}}
\end{figure*}
\par
In the left panel of the Fig.\ref{fig:each}, the IR8 ratios are plotted against the starburstiness for star-forming galaxies.
The AGN candidates are also shown for comparison.
The median value of the IR8 for the star-forming galaxies are constant at log($R_{\rm SB}$)$<$0 and increases with starburstiness at log($R_{\rm SB}$)$>$0.
The errors of the median values were defined as the 68th percentile of the data distribution divided by the square root of the number of samples in each bin.
Although the data show a non-negligible scatter compared with the local relation (black lines), our data are consistent with the results from Elbaz et al. (2011).
This result indicates the rise of IR8 with starburstiness among the star-forming galaxies in our sample, which suggests a relative ``weakness'' of PAH emission with respect to $L_{IR}$ for high $R_{\rm SB}$.
We refer to this weakness of PAH emission at high $R_{\rm SB}$ as a ``PAH deficit''.
We note that the median of the IR8 at lower $R_{\rm SB}$ may be overestimated due to the incompleteness of main-sequence galaxies, although it does not deny that high $R_{\rm SB}$ galaxies have a PAH deficit.
We note again that a fraction of our infrared luminosities are based only on L24 photometry, which does not affect our result.
This is because the redshift range of our sample is $z$$<$1.4, where $L_{IR}$ derived from 24$\mu$m photometry is consistent with those from far-infrared (Elbaz et al.2011), as shown in Fig.\ref{fig:lircom}.
\par
In the right panel, $\nu L(8)/\nu L(4.5)$ ratios are shown against starburstiness.
The median value for the star-forming galaxies increases with starburstiness at log($R_{\rm SB}$)$<$0.5 and stays constant at log($R_{\rm SB}$)$>$0.5, which is consistent again with the view that starburst galaxies have a relative ``weakness'' of PAH emission.
%with respect to $L_{IR}$.
On the other hand, the $\nu L(8)/\nu L(4.5)$ ratios of AGN candidates at higher $R_{\rm SB}$ are lower than those of star-forming galaxies.
Some objects show a higher luminosity ratio, log $\nu L(8)/\nu L(4.5)>1$, which is similar to the ``PAH selected galaxy'' of \cite{2010A&A...514A...5T}.
These galaxies are mainly distributed at moderate starburstiness, log $R_{\rm SB}$$\sim$-0.5-0.5.
We found that these galaxies have similar IR8 to other normal star-forming galaxies, which is consistent with \cite{2010A&A...514A...5T}.
Although some AGN candidates also show high $\nu L(8)/\nu L(4.5)$ ratio, this might be due to misclassification (see \S 2).
We note that a systematic bias could affect our result since we excluded AGN some of which might be star-forming (see \S \ref{data}).
However, since including these misclassified objects leads to lower $\nu L(8)/\nu L(4.5)$ ratios, as indicated in the right panel of Fig.\ref{fig:each}, it cannot affect our conclusion that starburst galaxies have a relative weakness of PAH emission.
\par
\begin{figure*}[ht]
\centering
\begin{minipage}{0.49\hsize}
\includegraphics[width=50mm,angle=270]{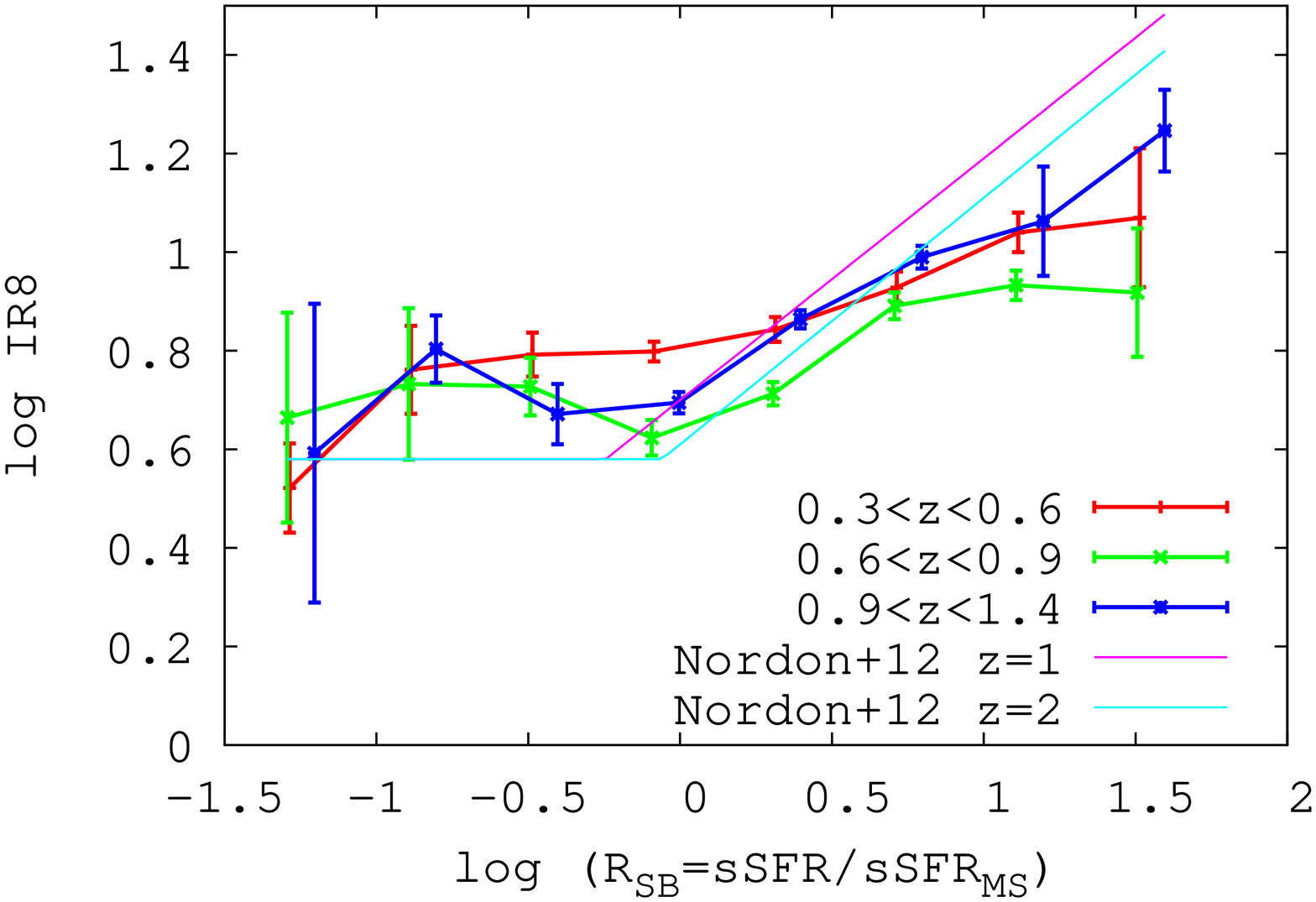}
\end{minipage}
\begin{minipage}{0.49\hsize}
\includegraphics[width=50mm,angle=270]{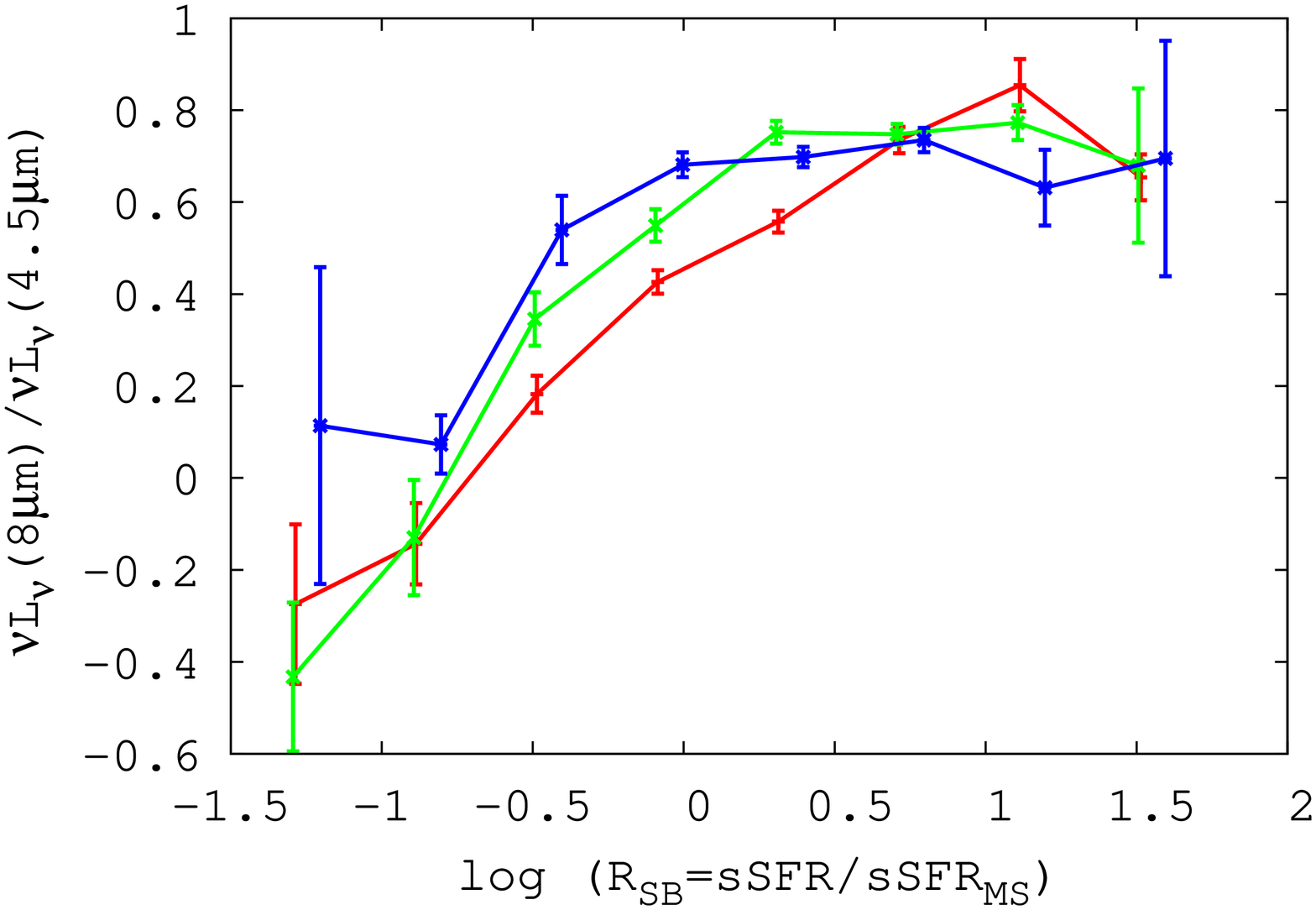}
\end{minipage}
\caption{
  Left: Redshift dependence of the relation between IR8 and $R_{\rm SB}$.
  Red, green, and blue lines show the median value of IR8 for star-forming galaxies at $z=$0.3-0.6, $z=$0.6-0.9, and $z=$0,9-1.4.
  Magenta and cyan lines show the relation for $z=$1 and $z=$2, which are determined by Nordon et al. (2012).
  Right: Same as left but for $\nu L(8)/\nu L(4.5)$ ratio.
\label{fig:def}}
\end{figure*}
\par
Lastly, we show the redshift dependence of the relation between PAH strength and starburstiness in Fig.\ref{fig:def}.
In the left panel, the median value of the IR8 ratios are shown against starburstiness with three redshift bins, $z=0.3-0.6$, $z=0.6-0.9$, and $z=0.9-1.4$.
For all redshift bins, IR8 ratios are constant at log($R_{\rm SB}$)$<$0 and increase with starburstiness at log($R_{\rm SB}$)$>$0.
This result is consistent with Nordon et al. (2012).
Although their curves show a redshift dependence, they mentioned the uncertainty of 0.1-0.15 dex in the methods to derive sSFR$_{MS}$, and $L_{IR}$ can account for the redshift dependence.
In the right panel, the median value of the $\nu L(8)/\nu L(4.5)$ ratios are shown against the starburstiness with the same redshift bins.
In all redshift bins, $\nu L(8)/\nu L(4.5)$ increases with starburstiness at log($R_{\rm SB}$)$<$0.5 and is constant at log($R_{\rm SB}$)$>$0.5.
\par
We consider here how the sample selection with the flux limit affects our results.
As mentioned in section \ref{data}, our MS is not complete, which leads a higher IR8 for lower $R_{\rm SB}$ galaxies.
Considering this effect is different in different redshift bins, we cannot reject a possibility that the similarity of the relation between IR8 and $R_{\rm SB}$ at different redshifts is only a coincidence.
Nonetheless, this effect cannot explain the higher IR8 for higher $R_{\rm SB}$, so that we can conclude that our results are qualitatively reliable.
%We also checked our sample selection does not affect our results.
On the other hand, the detection limits of 4.5$\mu$m and 8$\mu$m luminosities can lead to a lower and higher $\nu L(8)/\nu L(4.5)$ ratios, respectively.
To investigate this systematic effect, we applied flux cuts at various flux of S7 and S11 bands, which correspond to 4.5$\mu$m and 8$\mu$m at $z\sim$0.4.
Our results did not change in this test, confirming again our results are robust.

\section{Discussion}
\label{discuss}
Previous studies have established that the mid-infrared SED reflects the physical condition of galaxies rather than the infrared luminosity.
For example, Nordon et al. (2012) also showed that higher starburstiness galaxies have lower PAH emission and suggested that the dominant cause of the PAH weakness is the intense radiation fields by using the Spitzer space telescope.
However, they focused only on $z=1$ and $z=2$ galaxies owing to their wavelength coverage.
In our study, we explored the PAH behaviour of starburst galaxies at $z$=0.3-1.4, making use of the AKARI's continuous filter coverage at 2-24$\mu$m.
\par
In this section, we discuss the nature of the PAH deficit.
We refer to measures of the [CII] deficit as measured in the far-infrared, since the [CII] emission also originates in the PDRs and the deficit is seen at low and high-redshifts \cite[]{2003ApJ...594..758L,2009ApJ...701.1147A,2011ApJ...728L...7G,2013ApJ...774...68D,2013ApJ...776...38F}.
Here, we discuss five possibilities of the nature of the PAH deficit: a stronger continuum due to AGN or starburst, compact star-forming regions, a lack of ionising photons, low metallicity, and dust attenuation.
\par
If a strong continuum from the AGN or starburst dominate at mid-infrared luminosity, $\nu L(8)/\nu L(4.5)$ should be small.
However, we excluded all AGN candidates through the SED fitting with nine AKARI infrared bands, in which star-forming galaxies may be included.
Due to this conservative exclusion, the AGN are less likely to affect our results.
On the other hand, even with an extremely strong starformation with an ionisation parameter of $U=10^4$, the 4.5$\mu$m continuum cannot significantly be affected \cite[]{2007ApJ...657..810D}.
Hence, the cause of the PAH deficit is not likely to be the underlying continuum.
\par
If the PAHs are destroyed by strong UV radiation from the compact star-forming regions, the PAH emission weakens for a fixed L$_{IR}$ \cite[]{2004ApJ...613..986P}.
\cite{2011A&A...533A.119E} shows the IR8 correlates with star-formation compactness.
According to this relation, the PAH deficit galaxies have compact star-forming regions. 
It also predicts that galaxies with moderate starburstiness have extended star-forming regions.
At moderate starburstiness, log $R_{\rm SB}\sim 0$, in Fig.\ref{fig:each}, some galaxies with higher $\nu L(8)/\nu L(4.5)$ ratio were identified.
Takagi et al. (2010) argued that such high luminosity ratios cannot be explained unless they have an extended star-forming region.
Hence, these results support the possibility that PAHs are destroyed by strong UV radiation.
\par
%a lack of UV photons
On the other hand, if UV photons are absorbed by dust in HII regions and cannot reach the PAHs in PDRs, photoelectric heating of the PDRs should be less efficient, which results in both [CII] and PAH deficits \cite[]{2009ApJ...701.1147A,2011ApJ...728L...7G}.
We do not have enough data to confirm or reject this possibility.
%We have to note,
However, since a high ionisation parameter results in high dust absorption of UV photons \cite[]{2009ApJ...701.1147A}, we have to note that these two scenarios, ``a lack of UV photons'' and ``compact starburst'' could be connected.
\par
%metallicity
Galaxies with low metallicity should have a small amount of PAH particles, which results in lower PAH emission \cite[]{2005ApJ...628L..29E}.
To investigate this possibility, we compared the metallicity against the $\nu L(8)/\nu L(4.5)$ ratio.
%in Fig.\ref{fig:metal}.
The metallicity was derived from [NII]/$\rm H\alpha$ flux ratio observed with the Subaru/FMOS, as described in Appendix A.
We found no relation between metallicity and the luminosity ratio.
However, our metallicity range is only 12 + log O/H $>$ 8.5, although the PAH weakness can only be seen in 12 + log O/H$\ltsim$8.0 \cite[]{2005ApJ...628L..29E}.
Hence, we cannot strongly constrain this possibility.
%which is not wide enough, the metallicity is not likely to be the nature of the PAH deficit.
\par
%DUST ATTENUATION
Although dust attenuation is quite weak at mid-infrared wavelengths, it is still higher than in the far-infrared, so that IR8 can be higher due to dust attenuation.
However, if we assume 8$\mu$m and 4.5$\mu$m radiate from the same regions, dust attenuation should lead a higher $\nu L(8)/\nu L(4.5)$ ratio.
Furthermore, PAH deficit correlates with [CII] deficit, where dust attenuation is negligible. 
Hence, a constant $\nu L(8)/\nu L(4.5)$ indicates the intrinsic relative weakness of PAH emission in starburst galaxies.
\par
Summarising the discussions above, the dominant processes of the PAH deficit are probably due to a harsh radiation from compact starbursts and/or a lack of UV photons due to dust absorption.

\section{Conclusion}
We studied the PAH behaviour of galaxies at $z$=0.3-1.4 using 1868 galaxies sampled from the revised catalogue of the AKARI NEP-Deep survey.
The continuous wavelength coverage at 2-24$\mu$m of the AKARI/IRC enabled us to measure the $\nu L(8)/\nu L(4.5)$ ratio and the IR8 for the redshift range of interest.
All AGN candidates were separated with an SED fitting using nine AKARI filters.
We found the $\nu L(8)/\nu L(4.5)$ ratio increased at lower starburstiness, while it stayed constant at higher starburstiness.
Similarly, IR8 was constant at lower starburstiness and increased with starburstiness at higher starburstiness.
This relative weakness of the PAHs was seen throughout the redshift range $z=0.3-1.4$. 
We also found that galaxies with the highest $\nu L(8)/\nu L(4.5)$ ratio have moderate starburstiness.
These results can be interpreted as follows: starburst galaxies have compact star-forming regions, whose UV radiation destroys PAHs, and/or have dusty HII regions where a fraction of UV photons are absorbed by dust and PAHs cannot be excited.

\appendix
\begin{acknowledgements}
The AKARI NEP-Deep survey project activities are supported by JSPS grant 23244040. 
TM is supported by UNAM DGAPA Grant PAPIIT IN104113 and CONACyT Grant
179662.
\end{acknowledgements}
\section{Subaru/FMOS observation and data reduction}
\label{fmosobs}
The Subaru/FMOS \cite[]{2010PASJ...62.1135K} observations were conducted on 20-21 June 2012 to study star-forming galaxies at an intermediate redshift.
The 733 objects were selected with infrared luminosity, having $\rm H\alpha$ fluxes of 0.4$\times$10$^{-16}$erg cm$^{-2}$ s$^{-1}$ and were observed using a J-long band with Cross Beam Switch mode, where two fibres were allocated for each target.
One observes the target and another observes the sky $\sim$90 arcsec away from the target; the roles are switched at each exposure.
The fibre has a 1.2 arcsec aperture diameter, and the positional accuracy is 0.2 arcsec.
The on-source exposure time was 0.5-2h.
The spectral resolution of the J-long is $R$=1900.
\par
The data reduction was performed with the FMOS pipeline, FIBRE-pac \cite[]{2012PASJ...64...59I}.
The sky subtraction was conducted with A-B exposure for each observing set.
The wavelength calibration was carried out using OH masks, and the relative flux calibration was conducted with G-K type stars selected by the $g'r'i'$ colour-colour diagram.
Although fibre loss was not considered, which caused $\sim$0.2dex absolute flux error, it does not affect the line-flux ratio measurements.
\par
Among the 733 objects, 96 objects were found to have at least one emission line, of which 32 have both H$\alpha$ and [NII]$\lambda$6584 lines.
The line fluxes were measured by fitting with Gaussian functions.
The line ratios were converted to the metallicities using the relation \cite[]{2004MNRAS.348L..59P},
\begin{equation}
12 + \log(O/H)=8.90+0.57 \log([NII]\lambda 6584/H\alpha).
\end{equation}
Our targets have a metallicity range of 12 + log(O/H)=8.3-9.0.

\bibliographystyle{aa}
\bibliography{../BIB/ref,../BIB/fagn,../BIB/pah}

\begin{thebibliography}{42}
\expandafter\ifx\csname natexlab\endcsname\relax\def\natexlab#1{#1}\fi

\bibitem[{{Abel} {et~al.}(2009){Abel}, {Dudley}, {Fischer}, {Satyapal}, \& {van
  Hoof}}]{2009ApJ...701.1147A}
{Abel}, N.~P., {Dudley}, C., {Fischer}, J., {Satyapal}, S., \& {van Hoof},
  P.~A.~M. 2009, \apj, 701, 1147

\bibitem[{{Arnouts} {et~al.}(2007){Arnouts}, {Walcher}, {Le F{\`e}vre},
  {Zamorani}, {Ilbert}, {Le Brun}, {Pozzetti}, {Bardelli}, {Tresse}, {Zucca},
  {Charlot}, {Lamareille}, {McCracken}, {Bolzonella}, {Iovino}, {Lonsdale},
  {Polletta}, {Surace}, {Bottini}, {Garilli}, {Maccagni}, {Picat},
  {Scaramella}, {Scodeggio}, {Vettolani}, {Zanichelli}, {Adami}, {Cappi},
  {Ciliegi}, {Contini}, {de la Torre}, {Foucaud}, {Franzetti}, {Gavignaud},
  {Guzzo}, {Marano}, {Marinoni}, {Mazure}, {Meneux}, {Merighi}, {Paltani},
  {Pell{\`o}}, {Pollo}, {Radovich}, {Temporin}, \&
  {Vergani}}]{2007A&A...476..137A}
{Arnouts}, S., {Walcher}, C.~J., {Le F{\`e}vre}, O., {et~al.} 2007, \aap, 476,
  137

\bibitem[{{Bertin} \& {Arnouts}(1996)}]{1996A&AS..117..393B}
{Bertin}, E. \& {Arnouts}, S. 1996, \aaps, 117, 393

\bibitem[{{Bruzual} \& {Charlot}(2003)}]{2003MNRAS.344.1000B}
{Bruzual}, G. \& {Charlot}, S. 2003, \mnras, 344, 1000

\bibitem[{{Caputi} {et~al.}(2007){Caputi}, {Lagache}, {Yan}, {Dole},
  {Bavouzet}, {Le Floc'h}, {Choi}, {Helou}, \& {Reddy}}]{2007ApJ...660...97C}
{Caputi}, K.~I., {Lagache}, G., {Yan}, L., {et~al.} 2007, \apj, 660, 97

\bibitem[{{Chabrier}(2003)}]{2003PASP..115..763C}
{Chabrier}, G. 2003, \pasp, 115, 763

\bibitem[{{Chary} \& {Elbaz}(2001)}]{2001ApJ...556..562C}
{Chary}, R. \& {Elbaz}, D. 2001, \apj, 556, 562

\bibitem[{{Daddi} {et~al.}(2007){Daddi}, {Dickinson}, {Morrison}, {Chary},
  {Cimatti}, {Elbaz}, {Frayer}, {Renzini}, {Pope}, {Alexander}, {Bauer},
  {Giavalisco}, {Huynh}, {Kurk}, \& {Mignoli}}]{2007ApJ...670..156D}
{Daddi}, E., {Dickinson}, M., {Morrison}, G., {et~al.} 2007, \apj, 670, 156

\bibitem[{{D{\'{\i}}az-Santos} {et~al.}(2013){D{\'{\i}}az-Santos}, {Armus},
  {Charmandaris}, {Stierwalt}, {Murphy}, {Haan}, {Inami}, {Malhotra},
  {Meijerink}, {Stacey}, {Petric}, {Evans}, {Veilleux}, {van der Werf}, {Lord},
  {Lu}, {Howell}, {Appleton}, {Mazzarella}, {Surace}, {Xu}, {Schulz},
  {Sanders}, {Bridge}, {Chan}, {Frayer}, {Iwasawa}, {Melbourne}, \&
  {Sturm}}]{2013ApJ...774...68D}
{D{\'{\i}}az-Santos}, T., {Armus}, L., {Charmandaris}, V., {et~al.} 2013, \apj,
  774, 68

\bibitem[{{Draine} \& {Li}(2007)}]{2007ApJ...657..810D}
{Draine}, B.~T. \& {Li}, A. 2007, \apj, 657, 810

\bibitem[{{Elbaz} {et~al.}(2007){Elbaz}, {Daddi}, {Le Borgne}, {Dickinson},
  {Alexander}, {Chary}, {Starck}, {Brandt}, {Kitzbichler}, {MacDonald},
  {Nonino}, {Popesso}, {Stern}, \& {Vanzella}}]{2007A&A...468...33E}
{Elbaz}, D., {Daddi}, E., {Le Borgne}, D., {et~al.} 2007, \aap, 468, 33

\bibitem[{{Elbaz} {et~al.}(2011){Elbaz}, {Dickinson}, {Hwang},
  {D{\'{\i}}az-Santos}, {Magdis}, {Magnelli}, {Le Borgne}, {Galliano},
  {Pannella}, {Chanial}, {Armus}, {Charmandaris}, {Daddi}, {Aussel}, {Popesso},
  {Kartaltepe}, {Altieri}, {Valtchanov}, {Coia}, {Dannerbauer}, {Dasyra},
  {Leiton}, {Mazzarella}, {Alexander}, {Buat}, {Burgarella}, {Chary}, {Gilli},
  {Ivison}, {Juneau}, {Le Floc'h}, {Lutz}, {Morrison}, {Mullaney}, {Murphy},
  {Pope}, {Scott}, {Brodwin}, {Calzetti}, {Cesarsky}, {Charlot}, {Dole},
  {Eisenhardt}, {Ferguson}, {F{\"o}rster Schreiber}, {Frayer}, {Giavalisco},
  {Huynh}, {Koekemoer}, {Papovich}, {Reddy}, {Surace}, {Teplitz}, {Yun}, \&
  {Wilson}}]{2011A&A...533A.119E}
{Elbaz}, D., {Dickinson}, M., {Hwang}, H.~S., {et~al.} 2011, \aap, 533, A119

\bibitem[{{Elbaz} {et~al.}(2010){Elbaz}, {Hwang}, {Magnelli}, {Daddi},
  {Aussel}, {Altieri}, {Amblard}, {Andreani}, {Arumugam}, {Auld}, {Babbedge},
  {Berta}, {Blain}, {Bock}, {Bongiovanni}, {Boselli}, {Buat}, {Burgarella},
  {Castro-Rodriguez}, {Cava}, {Cepa}, {Chanial}, {Chary}, {Cimatti},
  {Clements}, {Conley}, {Conversi}, {Cooray}, {Dickinson}, {Dominguez},
  {Dowell}, {Dunlop}, {Dwek}, {Eales}, {Farrah}, {F{\"o}rster Schreiber},
  {Fox}, {Franceschini}, {Gear}, {Genzel}, {Glenn}, {Griffin}, {Gruppioni},
  {Halpern}, {Hatziminaoglou}, {Ibar}, {Isaak}, {Ivison}, {Lagache}, {Le
  Borgne}, {Le Floc'h}, {Levenson}, {Lu}, {Lutz}, {Madden}, {Maffei}, {Magdis},
  {Mainetti}, {Maiolino}, {Marchetti}, {Mortier}, {Nguyen}, {Nordon},
  {O'Halloran}, {Okumura}, {Oliver}, {Omont}, {Page}, {Panuzzo},
  {Papageorgiou}, {Pearson}, {Perez Fournon}, {P{\'e}rez Garc{\'{\i}}a},
  {Poglitsch}, {Pohlen}, {Popesso}, {Pozzi}, {Rawlings}, {Rigopoulou},
  {Riguccini}, {Rizzo}, {Rodighiero}, {Roseboom}, {Rowan-Robinson},
  {Saintonge}, {Sanchez Portal}, {Santini}, {Sauvage}, {Schulz}, {Scott},
  {Seymour}, {Shao}, {Shupe}, {Smith}, {Stevens}, {Sturm}, {Symeonidis},
  {Tacconi}, {Trichas}, {Tugwell}, {Vaccari}, {Valtchanov}, {Vieira},
  {Vigroux}, {Wang}, {Ward}, {Wright}, {Xu}, \& {Zemcov}}]{2010A&A...518L..29E}
{Elbaz}, D., {Hwang}, H.~S., {Magnelli}, B., {et~al.} 2010, \aap, 518, L29

\bibitem[{{Engelbracht} {et~al.}(2005){Engelbracht}, {Gordon}, {Rieke},
  {Werner}, {Dale}, \& {Latter}}]{2005ApJ...628L..29E}
{Engelbracht}, C.~W., {Gordon}, K.~D., {Rieke}, G.~H., {et~al.} 2005, \apjl,
  628, L29

\bibitem[{{Farrah} {et~al.}(2013){Farrah}, {Lebouteiller}, {Spoon},
  {Bernard-Salas}, {Pearson}, {Rigopoulou}, {Smith}, {Gonz{\'a}lez-Alfonso},
  {Clements}, {Efstathiou}, {Cormier}, {Afonso}, {Petty}, {Harris}, {Hurley},
  {Borys}, {Verma}, {Cooray}, \& {Salvatelli}}]{2013ApJ...776...38F}
{Farrah}, D., {Lebouteiller}, V., {Spoon}, H.~W.~W., {et~al.} 2013, \apj, 776,
  38

\bibitem[{{Graci{\'a}-Carpio} {et~al.}(2011){Graci{\'a}-Carpio}, {Sturm},
  {Hailey-Dunsheath}, {Fischer}, {Contursi}, {Poglitsch}, {Genzel},
  {Gonz{\'a}lez-Alfonso}, {Sternberg}, {Verma}, {Christopher}, {Davies},
  {Feuchtgruber}, {de Jong}, {Lutz}, \& {Tacconi}}]{2011ApJ...728L...7G}
{Graci{\'a}-Carpio}, J., {Sturm}, E., {Hailey-Dunsheath}, S., {et~al.} 2011,
  \apjl, 728, L7

\bibitem[{{Guo} {et~al.}(2013){Guo}, {Zheng}, \& {Fu}}]{2013ApJ...778...23G}
{Guo}, K., {Zheng}, X.~Z., \& {Fu}, H. 2013, \apj, 778, 23

\bibitem[{{Huang} {et~al.}(2009){Huang}, {Faber}, {Daddi}, {Laird}, {Lai},
  {Omont}, {Wu}, {Younger}, {Bundy}, {Cattaneo}, {Chapman}, {Conselice},
  {Dickinson}, {Egami}, {Fazio}, {Im}, {Koo}, {Le Floc'h}, {Papovich},
  {Rigopoulou}, {Smail}, {Song}, {Van de Werf}, {Webb}, {Willmer}, {Willner},
  \& {Yan}}]{2009ApJ...700..183H}
{Huang}, J.-S., {Faber}, S.~M., {Daddi}, E., {et~al.} 2009, \apj, 700, 183

\bibitem[{{Ilbert} {et~al.}(2006){Ilbert}, {Arnouts}, {McCracken},
  {Bolzonella}, {Bertin}, {Le F{\`e}vre}, {Mellier}, {Zamorani}, {Pell{\`o}},
  {Iovino}, {Tresse}, {Le Brun}, {Bottini}, {Garilli}, {Maccagni}, {Picat},
  {Scaramella}, {Scodeggio}, {Vettolani}, {Zanichelli}, {Adami}, {Bardelli},
  {Cappi}, {Charlot}, {Ciliegi}, {Contini}, {Cucciati}, {Foucaud}, {Franzetti},
  {Gavignaud}, {Guzzo}, {Marano}, {Marinoni}, {Mazure}, {Meneux}, {Merighi},
  {Paltani}, {Pollo}, {Pozzetti}, {Radovich}, {Zucca}, {Bondi}, {Bongiorno},
  {Busarello}, {de La Torre}, {Gregorini}, {Lamareille}, {Mathez}, {Merluzzi},
  {Ripepi}, {Rizzo}, \& {Vergani}}]{2006A&A...457..841I}
{Ilbert}, O., {Arnouts}, S., {McCracken}, H.~J., {et~al.} 2006, \aap, 457, 841

\bibitem[{{Iwamuro} {et~al.}(2012){Iwamuro}, {Moritani}, {Yabe}, {Sumiyoshi},
  {Kawate}, {Tamura}, {Akiyama}, {Kimura}, {Takato}, {Tait}, {Ohta}, {Totani},
  {Suzuki}, \& {Tonegawa}}]{2012PASJ...64...59I}
{Iwamuro}, F., {Moritani}, Y., {Yabe}, K., {et~al.} 2012, \pasj, 64, 59

\bibitem[{{Kimura} {et~al.}(2010){Kimura}, {Maihara}, {Iwamuro}, {Akiyama},
  {Tamura}, {Dalton}, {Takato}, {Tait}, {Ohta}, {Eto}, {Mochida}, {Elms},
  {Kawate}, {Kurakami}, {Moritani}, {Noumaru}, {Ohshima}, {Sumiyoshi}, {Yabe},
  {Brzeski}, {Farrell}, {Frost}, {Gillingham}, {Haynes}, {Moore}, {Muller},
  {Smedley}, {Smith}, {Bonfield}, {Brooks}, {Holmes}, {Curtis Lake}, {Lee},
  {Lewis}, {Froud}, {Tosh}, {Woodhouse}, {Blackburn}, {Content}, {Dipper},
  {Murray}, {Sharples}, \& {Robertson}}]{2010PASJ...62.1135K}
{Kimura}, M., {Maihara}, T., {Iwamuro}, F., {et~al.} 2010, \pasj, 62, 1135

\bibitem[{{Le Floc'h} {et~al.}(2005){Le Floc'h}, {Papovich}, {Dole}, {Bell},
  {Lagache}, {Rieke}, {Egami}, {P{\'e}rez-Gonz{\'a}lez}, {Alonso-Herrero},
  {Rieke}, {Blaylock}, {Engelbracht}, {Gordon}, {Hines}, {Misselt}, {Morrison},
  \& {Mould}}]{2005ApJ...632..169L}
{Le Floc'h}, E., {Papovich}, C., {Dole}, H., {et~al.} 2005, \apj, 632, 169

\bibitem[{{Luhman} {et~al.}(2003){Luhman}, {Satyapal}, {Fischer}, {Wolfire},
  {Sturm}, {Dudley}, {Lutz}, \& {Genzel}}]{2003ApJ...594..758L}
{Luhman}, M.~L., {Satyapal}, S., {Fischer}, J., {et~al.} 2003, \apj, 594, 758

\bibitem[{{Matsuhara} {et~al.}(2006){Matsuhara}, {Wada}, {Matsuura},
  {Nakagawa}, {Kawada}, {Ohyama}, {Pearson}, {Oyabu}, {Takagi}, {Serjeant},
  {White}, {Hanami}, {Watarai}, {Takeuchi}, {Kodama}, {Arimoto}, {Okamura},
  {Lee}, {Pak}, {Im}, {Lee}, {Kim}, {Jeong}, {Imai}, {Fujishiro}, {Shirahata},
  {Suzuki}, {Ihara}, \& {Sakon}}]{2006PASJ...58..673M}
{Matsuhara}, H., {Wada}, T., {Matsuura}, S., {et~al.} 2006, \pasj, 58, 673

\bibitem[{{Murakami} {et~al.}(2007){Murakami}, {Baba}, {Barthel}, {Clements},
  {Cohen}, {Doi}, {Enya}, {Figueredo}, {Fujishiro}, {Fujiwara}, {Fujiwara},
  {Garcia-Lario}, {Goto}, {Hasegawa}, {Hibi}, {Hirao}, {Hiromoto}, {Hong},
  {Imai}, {Ishigaki}, {Ishiguro}, {Ishihara}, {Ita}, {Jeong}, {Jeong},
  {Kaneda}, {Kataza}, {Kawada}, {Kawai}, {Kawamura}, {Kessler}, {Kester},
  {Kii}, {Kim}, {Kim}, {Kobayashi}, {Koo}, {Kwon}, {Lee}, {Lorente}, {Makiuti},
  {Matsuhara}, {Matsumoto}, {Matsuo}, {Matsuura}, {M{\"u}ller}, {Murakami},
  {Nagata}, {Nakagawa}, {Naoi}, {Narita}, {Noda}, {Oh}, {Ohnishi}, {Ohyama},
  {Okada}, {Okuda}, {Oliver}, {Onaka}, {Ootsubo}, {Oyabu}, {Pak}, {Park},
  {Pearson}, {Rowan-Robinson}, {Saito}, {Sakon}, {Salama}, {Sato}, {Savage},
  {Serjeant}, {Shibai}, {Shirahata}, {Sohn}, {Suzuki}, {Takagi}, {Takahashi},
  {Tanab{\'e}}, {Takeuchi}, {Takita}, {Thomson}, {Uemizu}, {Ueno}, {Usui},
  {Verdugo}, {Wada}, {Wang}, {Watabe}, {Watarai}, {White}, {Yamamura},
  {Yamauchi}, \& {Yasuda}}]{2007PASJ...59S.369M}
{Murakami}, H., {Baba}, H., {Barthel}, P., {et~al.} 2007, \pasj, 59, 369

\bibitem[{{Murata} {et~al.}(2013){Murata}, {Matsuhara}, {Wada}, {Arimatsu},
  {Oi}, {Takagi}, {Oyabu}, {Goto}, {Ohyama}, {Malkan}, {Pearson}, {Ma{\l}ek},
  \& {Solarz}}]{2013A&A...559A.132M}
{Murata}, K., {Matsuhara}, H., {Wada}, T., {et~al.} 2013, \aap, 559, A132

\bibitem[{{Noeske} {et~al.}(2007){Noeske}, {Weiner}, {Faber}, {Papovich},
  {Koo}, {Somerville}, {Bundy}, {Conselice}, {Newman}, {Schiminovich}, {Le
  Floc'h}, {Coil}, {Rieke}, {Lotz}, {Primack}, {Barmby}, {Cooper}, {Davis},
  {Ellis}, {Fazio}, {Guhathakurta}, {Huang}, {Kassin}, {Martin}, {Phillips},
  {Rich}, {Small}, {Willmer}, \& {Wilson}}]{2007ApJ...660L..43N}
{Noeske}, K.~G., {Weiner}, B.~J., {Faber}, S.~M., {et~al.} 2007, \apjl, 660,
  L43

\bibitem[{{Nordon} {et~al.}(2012){Nordon}, {Lutz}, {Genzel}, {Berta}, {Wuyts},
  {Magnelli}, {Altieri}, {Andreani}, {Aussel}, {Bongiovanni}, {Cepa},
  {Cimatti}, {Daddi}, {Fadda}, {F{\"o}rster Schreiber}, {Lagache}, {Maiolino},
  {P{\'e}rez Garc{\'{\i}}a}, {Poglitsch}, {Popesso}, {Pozzi}, {Rodighiero},
  {Rosario}, {Saintonge}, {Sanchez-Portal}, {Santini}, {Sturm}, {Tacconi},
  {Valtchanov}, \& {Yan}}]{2012ApJ...745..182N}
{Nordon}, R., {Lutz}, D., {Genzel}, R., {et~al.} 2012, \apj, 745, 182

\bibitem[{{Oi} {et~al.}(2014){Oi}, {Matsuhara}, {Murata}, {Goto}, {Wada},
  {Takagi}, {Ohyama}, {Malkan}, {Im}, {Shim}, {Serjeant}, \&
  {Pearson}}]{2014arXiv1403.7934O}
{Oi}, N., {Matsuhara}, H., {Murata}, K., {et~al.} 2014, ArXiv e-prints

\bibitem[{{Onaka} {et~al.}(2007){Onaka}, {Matsuhara}, {Wada}, {Fujishiro},
  {Fujiwara}, {Ishigaki}, {Ishihara}, {Ita}, {Kataza}, {Kim}, {Matsumoto},
  {Murakami}, {Ohyama}, {Oyabu}, {Sakon}, {Tanab{\'e}}, {Takagi}, {Uemizu},
  {Ueno}, {Usui}, {Watarai}, {Cohen}, {Enya}, {Ootsubo}, {Pearson}, {Takeyama},
  {Yamamuro}, \& {Ikeda}}]{2007PASJ...59S.401O}
{Onaka}, T., {Matsuhara}, H., {Wada}, T., {et~al.} 2007, \pasj, 59, 401

\bibitem[{{Peeters} {et~al.}(2004){Peeters}, {Spoon}, \&
  {Tielens}}]{2004ApJ...613..986P}
{Peeters}, E., {Spoon}, H.~W.~W., \& {Tielens}, A.~G.~G.~M. 2004, \apj, 613,
  986

\bibitem[{{P{\'e}rez-Gonz{\'a}lez} {et~al.}(2005){P{\'e}rez-Gonz{\'a}lez},
  {Rieke}, {Egami}, {Alonso-Herrero}, {Dole}, {Papovich}, {Blaylock}, {Jones},
  {Rieke}, {Rigby}, {Barmby}, {Fazio}, {Huang}, \&
  {Martin}}]{2005ApJ...630...82P}
{P{\'e}rez-Gonz{\'a}lez}, P.~G., {Rieke}, G.~H., {Egami}, E., {et~al.} 2005,
  \apj, 630, 82

\bibitem[{{Pettini} \& {Pagel}(2004)}]{2004MNRAS.348L..59P}
{Pettini}, M. \& {Pagel}, B.~E.~J. 2004, \mnras, 348, L59

\bibitem[{{Polletta} {et~al.}(2007){Polletta}, {Tajer}, {Maraschi},
  {Trinchieri}, {Lonsdale}, {Chiappetti}, {Andreon}, {Pierre}, {Le F{\`e}vre},
  {Zamorani}, {Maccagni}, {Garcet}, {Surdej}, {Franceschini}, {Alloin},
  {Shupe}, {Surace}, {Fang}, {Rowan-Robinson}, {Smith}, \&
  {Tresse}}]{2007ApJ...663...81P}
{Polletta}, M., {Tajer}, M., {Maraschi}, L., {et~al.} 2007, \apj, 663, 81

\bibitem[{{Polletta} {et~al.}(2006){Polletta}, {Wilkes}, {Siana}, {Lonsdale},
  {Kilgard}, {Smith}, {Kim}, {Owen}, {Efstathiou}, {Jarrett}, {Stacey},
  {Franceschini}, {Rowan-Robinson}, {Babbedge}, {Berta}, {Fang}, {Farrah},
  {Gonz{\'a}lez-Solares}, {Morrison}, {Surace}, \&
  {Shupe}}]{2006ApJ...642..673P}
{Polletta}, M.~d.~C., {Wilkes}, B.~J., {Siana}, B., {et~al.} 2006, \apj, 642,
  673

\bibitem[{{Rigby} {et~al.}(2008){Rigby}, {Marcillac}, {Egami}, {Rieke},
  {Richard}, {Kneib}, {Fadda}, {Willmer}, {Borys}, {van der Werf},
  {P{\'e}rez-Gonz{\'a}lez}, {Knudsen}, \& {Papovich}}]{2008ApJ...675..262R}
{Rigby}, J.~R., {Marcillac}, D., {Egami}, E., {et~al.} 2008, \apj, 675, 262

\bibitem[{{Salmi} {et~al.}(2012){Salmi}, {Daddi}, {Elbaz}, {Sargent},
  {Dickinson}, {Renzini}, {Bethermin}, \& {Le Borgne}}]{2012ApJ...754L..14S}
{Salmi}, F., {Daddi}, E., {Elbaz}, D., {et~al.} 2012, \apjl, 754, L14

\bibitem[{{Shim} {et~al.}(2013){Shim}, {Im}, {Ko}, {Jeon}, {Karouzos}, {Kim},
  {Lee}, {Papovich}, {Willmer}, \& {Weiner}}]{2013ApJS..207...37S}
{Shim}, H., {Im}, M., {Ko}, J., {et~al.} 2013, \apjs, 207, 37

\bibitem[{{Takagi} {et~al.}(2010){Takagi}, {Ohyama}, {Goto}, {Matsuhara},
  {Oyabu}, {Wada}, {Pearson}, {Lee}, {Im}, {Lee}, {Shim}, {Hanami}, {Ishigaki},
  {Imai}, {White}, {Serjeant}, \& {Malkan}}]{2010A&A...514A...5T}
{Takagi}, T., {Ohyama}, Y., {Goto}, T., {et~al.} 2010, \aap, 514, A5

\bibitem[{{Tielens}(2008)}]{2008ARA&A..46..289T}
{Tielens}, A.~G.~G.~M. 2008, \araa, 46, 289

\bibitem[{{Wada} {et~al.}(2008){Wada}, {Matsuhara}, {Oyabu}, {Takagi}, {Lee},
  {Im}, {Ohyama}, {Goto}, {Pearson}, {White}, {Serjeant}, {Wada}, \&
  {Hanami}}]{2008PASJ...60S.517W}
{Wada}, T., {Matsuhara}, H., {Oyabu}, S., {et~al.} 2008, \pasj, 60, 517

\bibitem[{{Werner} {et~al.}(2004){Werner}, {Roellig}, {Low}, {Rieke}, {Rieke},
  {Hoffmann}, {Young}, {Houck}, {Brandl}, {Fazio}, {Hora}, {Gehrz}, {Helou},
  {Soifer}, {Stauffer}, {Keene}, {Eisenhardt}, {Gallagher}, {Gautier}, {Irace},
  {Lawrence}, {Simmons}, {Van Cleve}, {Jura}, {Wright}, \&
  {Cruikshank}}]{2004ApJS..154....1W}
{Werner}, M.~W., {Roellig}, T.~L., {Low}, F.~J., {et~al.} 2004, \apjs, 154, 1

\end{thebibliography}
\end{document}